%% file: uprobencode.tex
\documentclass[english,onecolumn]{IEEEtran}
\usepackage[T1]{fontenc}
\usepackage[latin9]{inputenc}
\usepackage{color}
\usepackage{verbatim}
\usepackage{float}
\usepackage{amsmath}
\usepackage{amsthm}
\usepackage{amssymb}
\usepackage{esint}
\usepackage{graphicx}
\usepackage{psfrag}

\makeatletter
\theoremstyle{plain}
\newtheorem{thm}{\protect\theoremname}
\theoremstyle{definition}
\newtheorem{defn}{\protect\definitionname}
\theoremstyle{plain}
\newtheorem{lem}{\protect\lemmaname}
\theoremstyle{plain}
\newtheorem{cor}{\protect\corollaryname}
\theoremstyle{definition}
\newtheorem{example}{\protect\examplename}
\theoremstyle{definition}
\newtheorem{app}{\protect\applicationname}
\include{defns}

\usepackage{mathrsfs}

\usepackage{babel}
\providecommand{\corollaryname}{Corollary}
\providecommand{\definitionname}{Definition}
\providecommand{\lemmaname}{Lemma}
\providecommand{\theoremname}{Theorem}
\providecommand{\applicationname}{Application}
\providecommand{\examplename}{Example}

\makeatother

\begin{document}

\title{A Universal Coding Scheme for Remote Generation of Continuous Random Variables}

\author{Cheuk Ting Li and Abbas El Gamal}
\maketitle
\begin{abstract}

We consider a setup in which Alice selects a pdf $f$ from a set of prescribed pdfs $\Pr$ and sends a
prefix-free codeword $W$ to Bob in order to allow him to generate a single instance of the random variable $X\sim f$. We describe a universal coding scheme for this setup and establish an upper bound on the expected codeword length when the pdf $f$ is bounded, orthogonally concave (which includes quasiconcave pdf), and has a finite first absolute moment. A dyadic decomposition scheme is used to express the pdf as a mixture of uniform pdfs over hypercubes. Alice randomly selects a hypercube according to its weight, encodes its position and size into $W$, and sends it to Bob who generates $X$ uniformly over the hypercube.
Compared to previous results on channel simulation, our coding scheme applies to any continuous distribution and does not require two-way communication or shared randomness. We apply our coding scheme to classical simulation of quantum entanglement and obtain a better bound on the average codeword length than previously known.

%We introduce the notion of universal distribution code, which is a
%code taking a pdf $f_{X}$ as input, and outputs a prefix-free codeword
%$W$ that can be used to generate one random variate $X\sim f_{X}$
%without any other knowledge on $f_{X}$. We construct a code which
%can accept any orthogonally concave $f_{X}$ (which includes quasiconcave
%pdf) that is bounded (i.e., $\sup_{x}f_{X}(x)<\infty$) and has finite
%first moment $\E(\left\Vert X\right\Vert _{\infty})<\infty$,
%and present a bound on the expected codeword length in terms of $\sup_{x}f_{X}(x)$
%and $\E(\left\Vert X\right\Vert _{\infty})$. Applications
%of the universal distribution code includes distributed channel synthesis
%without a fixed input distribution, distributed simulation of random
%variables as in Wyner's common information, and lossy compression
%with a fixed distribution of the error.

\end{abstract}


\section{Introduction}

\begin{figure}[h!]
\begin{center}
\psfrag{y}[b]{$W\in \{0,1\}^*$} 
\psfrag{p}[c]{Encoder} 
\psfrag{d}[c]{Decoder}
\psfrag{x}[r]{$p\in \Pr$} 
\psfrag{mh}[l]{$X\sim p$} 
\includegraphics[scale=0.55]{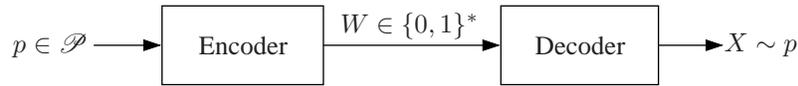}
\caption{Universal remote generation of random variables.}
\label{fig:setting} 
\end{center}
\end{figure}

Consider the one-shot remote random variable generation setting depicted in Figure~\ref{fig:setting}. Alice and Bob
both agree on a set of distributions $\Pr$ (over a discrete or continuous set). Alice selects a distribution
$p\in\Pr$ and wishes to have Bob generate a random variable $X$
according to this distribution. To accomplish this goal, Alice and Bob use an agreed upon \emph{universal} coding scheme in which Alice uses a stochastic encoder to assign to each $p\in \Pr$ a codeword $W\in \{0,1\}^*$ from an agreed upon prefix-free code and Bob uses a stochastic decoder to generate a single instance of $X\sim p$ from the received codeword $W$. Let $L(W)$ be the length of $W$ in bits. Is there a coding scheme such that for every distribution $p \in \Pr$, Bob can generate $X \sim p$ with finite expected codeword length $\E_p (L(W))$?

The answer to this question clearly depends on the set of distributions $\Pr$. Consider the following two simple special cases:
\begin{itemize}
\item[1.] Let $\Pr$ be the set of probability mass functions over the integers,
then we can use the following ``generate--compress'' strategy. Alice generates $X\sim p$ and then uses a universal code over the integers, e.g., \cite{levenshtein1968redundancy,elias1975}, 
to encode $X$ into $W$. Upon receiving $W$, Bob recovers $X$. Using these codes, the expected codeword length  $\E_p (L(W))$ is finite as long as $\E_p (\log X)$ is finite.
Note that this scheme uses a stochastic encoder but a deterministic decoder.  

\item[2.] Let $X$ be continuous and the class of pdfs $\Pr$
has a finite (or countable) cardinality, then we can use the following ``compress--generate'' strategy. Alice encodes the index of
$p$ into $W$. Upon receiving $W$, Bob first recovers $p$
then use it to generates $X$. Note that in this scheme, the encoder is deterministic but the decoder is stochastic.
\end{itemize}

If we index the set $\Pr$ by $\theta \in \Theta$, then our setting can be viewed as a one-shot synthesis (or simulation) of a channel from $\theta$ to $X$ with only one-way communication and without common randomness. Several channel simulation scenarios have been previously studied in classical and quantum information theory. In~\cite{bennett2002entanglement}, Bennett et al. considered the asymptotic setting and established the reverse Shannon theorem, which states that $k$ uses of a channel with capacity $C$ can be simulated using $k C+o(k)$ bits of communication with unlimited amount of common randomness. In~\cite{winter2002compression}, Winter studied the asymptotic case with limited common randomness and $\theta_i$ distributed according to a given distribution. He showed that $k I(\theta ; X) + o(k)$ bits of communication and $k H(X|\theta)+o(k)$ bits of common randomness suffice. Subsequently, Cuff~\cite{cuff2013distributed} characterized the entire tradeoff region between communication and common randomness for the same setting.

For the one-shot channel simulation setting, schemes based on rejection sampling were developed  by Steiner~\cite{steiner2000towards}, who assumed that Alice and Bob share unlimited common randomness, and by Massar et al.~\cite{massar2001classical}, who assumed two-way communication between Alice and Bob. Harsha et al.~\cite{harsha2010communication} established a one-shot version of the reverse Shannon theorem using rejection sampling.  These rejection sampling schemes, however, are sensitive to the size of $\Pr$\,---\,a large size $\Pr$ leads to a high rejection rate, which in turn leads to a high computation time.

Note that the aforementioned asymptotic and one-shot channel simulation schemes are not universal since a scheme designed for a channel from $\theta$ to $X$ is guaranteed to work only when the simulated distribution lies in the convex hull of the set of output distributions $\{p(x|\theta)\}$. 

In this paper, we present a universal coding scheme for remote generation of continuous random variables (over scalars or vectors), which we will refer to as \emph{universal dyadic coding scheme}. When $\Pr$ is restricted to the set of orthogonally concave pdfs $\{f_X(x)\}$, (which includes quasiconcave),
we are able to establish an upper bound on the expected codeword length of $W$ in terms of $\sup_{x}f_{X}(x)$ and $\E(\left\Vert X\right\Vert _{\infty})$.
Our scheme uses a dyadic decomposition to express the selected pdf as a mixture of uniform distributions over hypercubes. Alice first selects a hypercube from this mixture at random according to its weight, then encodes its position and size into a codeword $W$ using an agreed upon universal
code over the integers. Upon receiving $W$, Bob finds the hypercube and generates $X$ uniformly over it.

In~\cite{distsimcont_arxiv}, a similar dyadic decomposition scheme was introduced for distributed simulation of continuous random variables according to an agreed upon pdf in a non-universal manner. In Section~\ref{sec:unif} we provide a more detailed comparison between these two dyadic coding schemes.

To further motivate our setup and universal dyadic coding scheme, consider the following two applications. 
\smallskip

\begin{app}[Classical simulation of quantum entanglement] \label{ex:simquantum}
%\noindent 1. {\bf Classical simulation of quantum entanglement.}

The simulation of correlations induced by quantum entanglement using classical communication has been widely studied, e.g., see~\cite{maudlin1992bell,brassard1999cost,steiner2000towards}.
Consider the Bell state $|\Phi^+\rangle=(|0\rangle_A |0\rangle_B+|1\rangle_A|1\rangle_B)/\sqrt{2}$ of a pair of qubits~\cite{bell1964einstein}, one held by Alice and the other held by Bob.
If Alice measures her qubit in the direction $\theta_A$ (unknown to Bob) to obtain $Y_A\in \{\pm 1\}$ and Bob measures his qubit in the direction $\theta_B$ (unknown to Alice) to obtain $Y_B\in \{\pm 1\}$, then $\P\{Y_A=1\}=\P\{Y_B=1\}=1/2$ and $\E[Y_A Y_B]=-\cos(\theta_A-\theta_B)$.
By Bell's theorem, it is impossible to simulate the joint distribution of $(Y_A,Y_B)$ for all $\theta_A$ and $\theta_B$ using a classical common randomness source (local hidden variables) between Alice and Bob in place of the qubits. However, such simulation is possible if we instead allow Alice to send a codeword $W$ to Bob.
By a modification of the expression in~\cite{feldmann1995new} and letting $X\in [0, 2\pi]$ be a random variable with conditional pdf
\begin{equation}
f(x|y_{A};\theta_{A})=\frac{1}{2}\max\{\cos(y_{A}(x-\theta_{A})),\,0\},\label{eq:bell}
\end{equation}
and $Y_B = -\sgn (\cos(X-\theta_B))$, then $(Y_A,Y_B)$ follows the desired distribution.
Hence Alice can generate $Y_A$ and use our universal remote generation coding scheme to encode $f(x|Y_A;\theta_A)$ into $W$ to allow Bob to generate $X$ and $Y_B$. Using Theorem~\ref{thm:bdd_len} in Section~\ref{sec:bounded}, we show that the expected number of bits is bounded as $\E(L(W)) \le 12.31$, and using numerical computation we show that $\E(L(W)) \le 8.96$ is achievable. In comparison, the scheme in~\cite{massar2001classical}, which requires two-way communication (we only allow one-way) provides a looser upper bound of 20 bits on the average number of bits needed. 
\end{app}

\smallskip

\begin{app}[Minimax mixed strategy with a helper] \label{ex:minimax}
% \noindent 2. {\bf Minimax mixed strategy with a helper.} 

In decision theory, it is sometimes desirable to adopt a mixed strategy in which
the decision is chosen at random. Suppose the payoff $g(X,\theta)$ depends
on the agent's decision $X$ and an unknown parameter $\theta$ selected from a set $\Theta$ (which may be chosen by an adversary). The optimal minimax mixed strategy to choose $X$ is 
\[
F_{X}^{*}(x)=\underset{F_{X}(x)}{\arg\max}\inf_{\theta \in \Theta}\E \left[g(X,\theta)\right].
\]
Now suppose the parameter $\theta=(\theta_{1},\theta_{2})$, where $\theta_{1}$ is known to a helper
(Alice) but $\theta_{2}$ is not known to Alice or the decision agent (Bob). Alice wishes to help Bob generate the decision $X$ with the optimal pdf given $\theta_1$,
\[
F_{X}^{*}(x;\theta_{1})=\underset{F_{X}(x)}{\arg\max}\inf_{\theta_{2}}\E\left[g(X,\theta_{1},\theta_{2})\right].
\]
To help Bob generate $X$ using this optimal pdf, Alice can use our universal remote generation coding scheme.
For example, consider the payoff function
\[
g(x,\theta)=\begin{cases}
e^{2\theta-x} & \text{if}\;x\ge\theta\\
0 & \text{if}\;x<\theta,
\end{cases}
\]
where $\theta\ge 0$. If nothing else is known
about $\theta$, then the optimal minimax mixed strategy would be to
choose $X$ according to the pdf $f_{X}^{*}(x)=e^{-x}$ for $x\ge0$, which guarantees a payoff of $1/2$.
Now assume that Alice knows that $\theta \ge a \ge 0$, then the optimal mixed strategy is  $f_{X}^{*}(x;a)=e^{-(x-a)}$
for $x\ge a${, which results in a payoff of $(1/2)e^a$}. As shown in Theorem~\ref{thm:nonunif_len} in Section \ref{sec:nonunif}, Alice can use our universal remote generation coding scheme to enable Bob to generate $X$ according to this pdf using no more than $\log(a+1) + 2 \log(\log(a+1) + 12) + 23$ bits on average. Our universal coding scheme can also be used to perform mixed strategies in other scenarios, e.g., Nash equilibrium~\cite{nash1951non} in non-cooperative games.
\end{app}

The rest of the paper is organized as follows. For clarity of presentation, we first present the construction of our universal dyadic coding scheme for uniform distributions over subsets of $\mathbb{R}^n$ and upper bound its expected codeword length. In Section \ref{sec:nonunif}, we extend our scheme to non-uniform distributions, establishing an upper bound on 
 the expected codeword length for orthogonally concave pdfs. 
In Section \ref{sec:bounded}, we present a variant of our scheme for distributions with a uniform bounded support and apply it to the simulation of the Bell state. Finally, in Section \ref{sec:lbound}, we present a lower bound on the expected codeword length in terms of the relative entropy between the actual and the implied distribution of our scheme.

\subsection{Notation}

Throughout this paper, we assume that $\log$ is base 2 and entropy
$H$ is in bits. Log in base $e$ is written as $\ln$. We use the
notation: $[a:b]=[a,b]\cap\mathbb{Z}$,

%Denote the length of $w\in\left\{ 0,1\right\} ^{*}$ as $L(w)$.

A set $A\subseteq\mathbb{R}^{n}$ is orthogonally convex if for any
line $L$ parallel to one of the $n$ axes, $L\cap A$ is a connected
set (empty, a point or an interval). A function $f$ is orthogonally
concave if the hypograph $\left\{ (x,\alpha):\,x\in\mathbb{R}^{n},\,\alpha\le f(x)\right\} $
is orthogonally convex.

For a Lebesgue measurable set $A\subseteq\mathbb{R}^{n}$,
we define the volume $\mathrm{V}_{n}(A)=\int_{\mathbb{R}^{n}}\mathbf{1}_{A}(x)dx$.
For $A,B\subseteq\mathbb{R}^{n}$, $A+B$ denotes the Minkowski sum
$\left\{ a+b:\,a\in A,\,b\in B\right\} $, and for $x\in\mathbb{R}^{n}$,
$A+x=\left\{ a+x:\,a\in A\right\} $. For $\gamma\in\mathbb{R}$,
$\gamma A=\left\{ \gamma a:\,a\in A\right\} $. For $M\in\mathbb{R}^{n\times n}$,
$MA=\left\{ Ma:\,a\in A\right\} $. The \emph{erosion} $A\ominus B$
is defined as $\left\{ x\in\mathbb{R}^{n}:\,B+x\subseteq A\right\} $.

\section{Uniform Distributions\label{sec:unif}}

In this section, we develop our universal dyadic coding scheme for the set of uniform pdfs over finite volume sets $A\subseteq \mathbb{R}^n$. We first introduce the dyadic decomposition of a set~\cite{distsimcont_arxiv}, which is the building
block of our coding scheme. 
\begin{defn}[Dyadic decomposition]
For $v\in\mathbb{Z}^{n}$ and $k\in\mathbb{Z}$, define the hypercube
$C_{k,v}=2^{-k}([0,1]^{n}+v)\subset\mathbb{R}^{n}$. For a set
$A\subseteq\mathbb{R}^{n}$ with a boundary of measure zero and $k\in\mathbb{Z}$, define the set
\[
D_{k}(A)=\left\{ v\in\mathbb{Z}^{n}\,:\,C_{k,v}\subseteq A\text{ and }C_{k-1,\left\lfloor v/2\right\rfloor }\nsubseteq A\right\} ,
\]
where $\left\lfloor v/2\right\rfloor $ is the vector formed by the
entries $\left\lfloor v_{i}/2\right\rfloor $. The \emph{dyadic decomposition}
of $A$ is the partitioning of $A$ into hypercubes $\left\{ C_{k,v}\right\} $
such that $v\in D_{k}(A)$ and $k\in\mathbb{Z}$.
Since every point $x$ in the interior of $A$ is contained in some hypercube in $A$, the interior of $A$ is contained in 
 $\cup_{k\in\mathbb{Z},\,v\in D_{k}(A)}C_{k,v}$, and the set of points in $A$ not covered by the hypercubes has measure zero.
\end{defn}

Our scheme uses a universal code over the integers to encode the position and size of the hypercubes. In particular, we will use the signed Elias delta code defined as follows~\cite{elias1975}. Let
%Recall that the Elias gamma code $g_{\gamma+}: \{1,2,\ldots\} \to \{0,1\}^*$, unsigned Elias delta code $g_{\delta+}: \{1,2,\ldots\} \to \{0,1\}^*$ and  are defined as follows.
\begin{align*}
g_{\gamma+}(k) &= 0^N \,\Vert\, 1 \,\Vert\, a_{N-1} a_{N-2} \ldots a_0,\\
g_{\delta+}(k) &= g_{\gamma+}(N+1) \,\Vert\, a_{N-1} a_{N-2} \ldots a_0.
\end{align*}
Then the signed Elias code is
\[
g_{\delta}(k) = \begin{cases}
g_{\delta+}(1-2k) &\text{if}\; k \le 0,\\
g_{\delta+}(2k) &\text{if}\; k > 0,
\end{cases}
\]
where $a_{N} a_{N-1} \ldots a_0$ is the binary representation of $k$. The length of the codeword $g_{\delta}(k)$ is
\begin{equation}
\begin{aligned}L(g_{\delta}(k)) & =\left\lfloor \log(2|k|+1)\right\rfloor +2\left\lfloor \log\left(\left\lfloor \log(2|k|+1)\right\rfloor +1\right)\right\rfloor +1.\end{aligned}
\label{eq:len_delta}
\end{equation}

We are now ready to define the universal dyadic coding scheme for the set of uniform pdfs.

\begin{figure}[h!]
\begin{center}
\small
\psfrag{a}[c]{$101011$} 
\psfrag{b}[c]{$111$} 
\psfrag{c}[c]{$101010101$}
\psfrag{d}[c]{$110101$} 
\psfrag{a1}[c]{$010010100$} 
\psfrag{b1}[c]{$010001000100$} 
\psfrag{c1}[c]{$010011$}
\psfrag{d1}[c]{$010001001$} 
\psfrag{a2}[c]{\tiny $011000110001110$} 
\psfrag{b2}[c]{\tiny $011000111001110$} 
\psfrag{c2}[c]{\tiny $011000110001100$}
\psfrag{d2}[c]{\tiny $011000111001100$} 
\psfrag{e}[t]{$0, (-1,0)$} 
\psfrag{f}[t]{$0, (0,0)$} 
\psfrag{g}[t]{$0, (-1,-1)$}
\psfrag{h}[t]{$0, (0,-1)$} 
\psfrag{e1}[t]{$1, (0,1)$} 
\psfrag{f1}[t]{$1, (1,1)$} 
\psfrag{g1}[t]{$1, (0,0)$}
\psfrag{h1}[t]{$1, (1,0)$} 
\psfrag{e2}[t]{$2, (2,3)$} 
\psfrag{f2}[t]{$2, (3,3)$} 
\psfrag{g2}[t]{$2, (2,2)$}
\psfrag{h2}[t]{$2, (3,2)$} 
\includegraphics[scale=0.83]{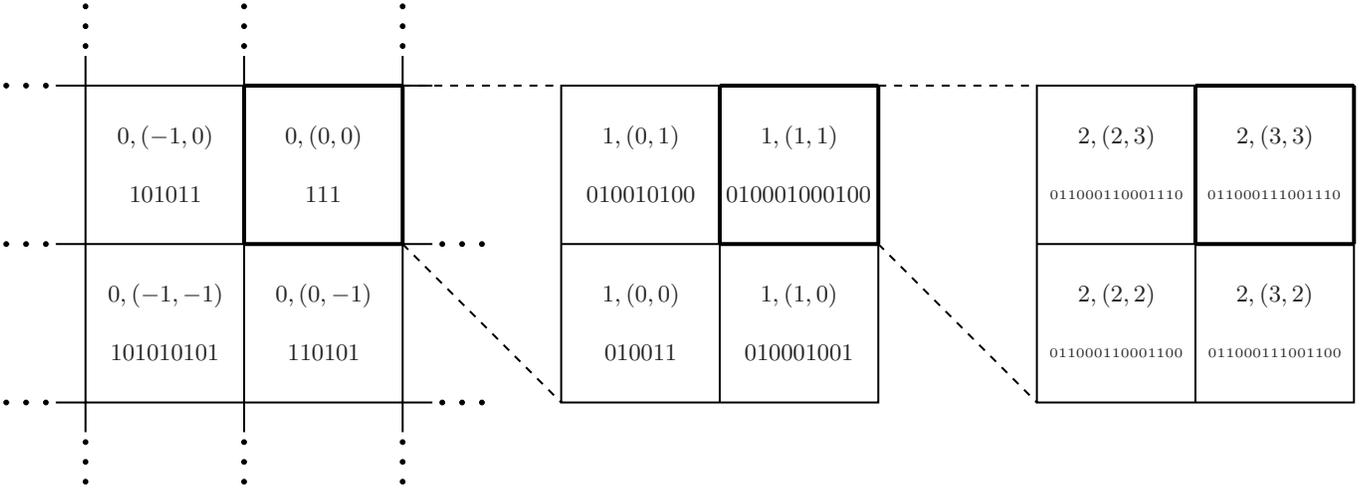}
\caption{Dyadic squares $C_{k,v}$ used in the dyadic decomposition for $n=2$, showing their associated $k$, $v$ and codeword assignments.}
\label{fig:dyadic} 
\end{center}
\end{figure}

\noindent {\bf Universal dyadic coding scheme for uniform pdfs}.
The universal dyadic coding scheme for the set of uniform pdfs over positive, finite volume subsets $A \subset \mathbb{R}^n$ with a boundary of measure zero consists of:
\begin{enumerate}
\item A stochastic encoder that generates $\xt$ according to the observed uniform pdf over $A$. It then finds $w=(k,v)$ such that $v\in D_{k}(A)$ and $\xt\in C_{k,v}$. The encoder then maps $(k,v)$ into a codeword $w$ which consists of the concatenation of
signed Elias delta codewords  for $k,v_{1},\ldots,v_{n}$, i.e.,
%The proposed code utilizes the signed Elias delta code $g_{\delta}(i) \in \{0,1\}^*$
$w=g_{C}(k,v)=g_{\delta}(k)\,\Vert\, g_{\delta}(v_{1})\,\Vert\,\cdots\,\Vert\, g_{\delta}(v_{n})$.
\item A stochastic decoder that upon receiving $w$ recovers $(v,k)$ and generates $x$ according to a uniform pdf over $C_{k,v}$.
\end{enumerate}
The dyadic decomposition for $\mathbb{R}^2$ and the assignments of codeword to the squares are illustrated in Figure~\ref{fig:dyadic}.

The following illustrates how our scheme is used for a given pdf.
\begin{example}
Consider a uniform pdf over the ellipse $A=\left\{ x\in\mathbb{R}^{2}:\, x^{T}Kx<1\right\} $, $K=\left[\begin{array}{cc}
4/3 & -2/3\\
-2/3 & 4/3
\end{array}\right]$. Figure~\ref{fig:unif_ellipse} depicts the universal dyadic coding scheme for this pdf. The encoder first generates a point in the ellipse uniformly at random, and then sends the codeword representing the square containing the point. The expected codeword length (computed by listing all squares in the dyadic decomposition with side length at least $2^{-16}$) is $15.6$. Note that the entropy of $W$, $H(W) = 6.35$ is significantly smaller since the code is universal.
\begin{figure}[h!]
\begin{center}
\includegraphics[scale=0.3]{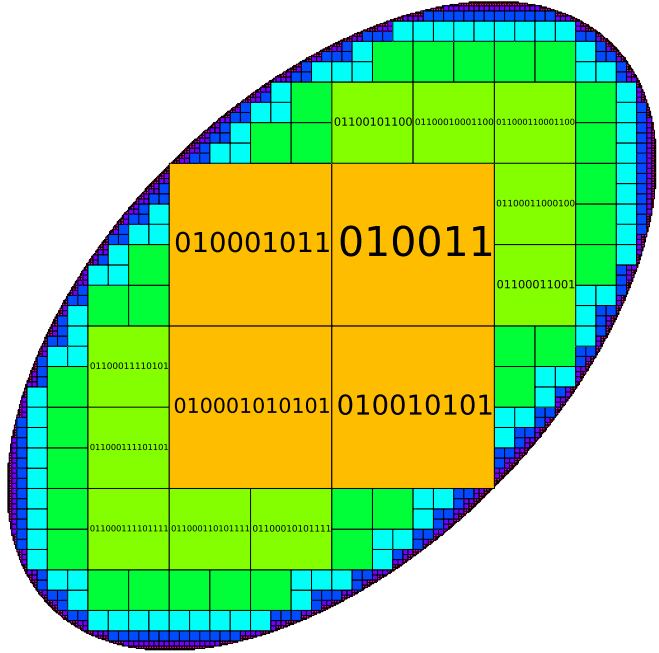}
% \caption{Top left, top right and bottom left: The codewords assigned to the dyadic squares with side lengths $1$, $1/2$ and $1/4$ respectively in universal dyadic coding scheme for $n=2$. Bottom right: Universal dyadic coding scheme on the uniform distribution on an ellipse.}
\caption{Universal dyadic coding scheme on the uniform distribution over an ellipse.}
\label{fig:unif_ellipse} 
\end{center}
\end{figure}
\end{example}

%\begin{figure}[H]
%\[
%F_{X^{n}}\longrightarrow\boxed{\text{Enc}}\stackrel{W}{\longrightarrow}\begin{array}{ccc}
%\boxed{\text{Dec 1}} & \longrightarrow & X_{1}\\
%\vdots &  & \vdots\\
%\boxed{\text{Dec \ensuremath{n}}} & \longrightarrow & X_{n}
%\end{array}
%\]
%
%
%\caption{Universal distribution code for distributed generation}
%\end{figure}

%To bound the expected codeword length for the universal dyadic coding scheme, recall that 
%recall the definition of the erosion entropy in~\cite{distsimcont_arxiv}.

The length of the codeword of the universal dyadic coding scheme depends on the magnitude of $k$ and $v_1,\ldots,v_n$, (which depends on $k$ and $\Vert x \Vert$), hence the length can be bounded using $k$ and $\Vert x \Vert$. in~\cite{distsimcont_arxiv}, it is shown that the expected value of $k$ can be bounded using the following quantity.

\begin{defn}[Erosion entropy]
The \emph{erosion entropy} of the set $A$ by the set $B$, where
$A\subseteq\mathbb{R}^{n}$ with $\mathrm{V}_{n}(A)<\infty$, and
$B\subseteq\mathbb{R}^{n}$ is a convex set, is defined as 
\[
h_{\ominus B}(A)=\int_{-\infty}^{\infty}\left(\mathbf{1}\left\{ t\ge0\right\} -\frac{\mathrm{V}_{n}\left(A\ominus2^{-t}B\right)}{\mathrm{V}_{n}(A)}\right)dt,
\]
where $A\ominus B=\left\{ x\in\mathbb{R}^{n}:\,B+x\subseteq A\right\} $
is the erosion of $A$ by $B$. 
\end{defn}
If $A$ is orthogonally convex, the erosion entropy of $A$ by the
hypercube $[0,1]^{n}$ can be bounded by the expected norm of
the uniform distribution on $A$, as shown in the following. 
\begin{lem}
\label{lem:erosion_bd_ortho_norm}Let the set $A\subseteq\mathbb{R}^{n}$ be orthogonally convex
with $\mathrm{V}_{n}(A)<\infty$, and let $X\sim\U(A)$, then
\textnormal{
\[
h_{\ominus[0,1]^{n}}(A)\le(n-1)\log\E\left[\left\Vert X\right\Vert _{\infty}\right]-\log\mathrm{V}_{n}(A)+4n.
\]
}
\end{lem}
The proof of this lemma is given in 
Appendix~\ref{apx:erosion_bd_ortho_norm}. We now use the erosion entropy to bound the expected codeword length
of the universal dyadic coding scheme. 
\begin{thm}
\label{thm:unif_len}
The expected codeword length of the universal dyadic coding scheme for uniform pdfs
for $X \sim \U(A)$ is upper bounded as
% Let $\Pr$ be the set of uniform pdfs over sets $A\subseteq\mathbb{R}^{n}$, that is, $X\sim\U(A)$ for some $A$. Then, the expected codeword length of the universal dyadic coding scheme is upper bounded as 
\textnormal{
\[
\begin{aligned}\E\left[L(W)\right] & \le n \ell_{\delta}\left(h+\E\left[ \log\left(\left\Vert X\right\Vert _{\infty}+\mathrm{V}_{n}^{1/n}(A)\right)\right]+4\right)+\ell_{\delta}\left(\log\left(h+2\max\left\{ \log\mathrm{V}_{n}^{1/n}(A),\,0\right\} +\frac{5}{2}\right)+2\right)\\
 & \le n \ell_{\delta}\left(h+\log\E[\left\Vert X\right\Vert _{\infty}]+8\right)+\ell_{\delta}\left(\log\left(h+2\max\left\{ \log\E[\left\Vert X\right\Vert _{\infty}],\,0\right\} +9\right)+2\right),
\end{aligned}
\]
}
where $\ell_{\delta}(t)=t+2\log t$ and $h=h_{\ominus[0,1]^{n}}(A)$. 
\end{thm}
Theorem~\ref{thm:unif_len} shows that the expected codeword length
depends on the erosion entropy, the expected magnitude of $X$,
and the volume of the set. Intuitively, the erosion entropy measures
the complexity of the set (or loosely speaking its surface area to volume
ratio). However, the erosion entropy is invariant under shifting. Since our universal scheme
is sensitive to the position of $A$ as well its shape, the bound in Theorem~\ref{thm:unif_len}
depends also on the expected magnitude of $X$. The function
$\ell_{\delta}(t)$ in Theorem~\ref{thm:unif_len} comes from the length of the Elias
delta code in \eqref{eq:len_delta}. Other universal codes for integers
may be used in place of Elias delta code, and result in a different
bound.

We now prove Theorem \ref{thm:unif_len}.
\begin{IEEEproof}[Proof of Theorem \ref{thm:unif_len}]
Let $x\in A$. Consider the length of the codeword for $(v,k)$ with
$v\in D_{k}(A)$ and $x\in C_{k,v}$. We have $x\in2^{-k}\left([0,1]^{n}+v\right)$,
hence $\left\Vert x\right\Vert _{\infty}\ge2^{-k}\max_{i}\left(\left|v_{i}+1/2\right|-1/2\right)$.
Since $2^{-nk}\le\mathrm{V}_{n}(A)$, $k\ge-(1/n)\log\mathrm{V}_{n}(A)$.
Let 
\[
\tau=2\max\left\{ (1/n)\log\mathrm{V}_{n}(A),\,0\right\} ,
\]
then $|k|\le k+\tau$. From (\ref{eq:len_delta}), the length of the
codeword for $(v,k)$ is 
\[
\begin{aligned}  L(g_{C}(v,k))
 & \le\left\lfloor \log(2|k|+1)\right\rfloor +2\left\lfloor \log\left(\left\lfloor \log(2|k|+1)\right\rfloor +1\right)\right\rfloor \\
 & \;\;\;\;+\sum_{i=1}^{n}\left(\left\lfloor \log(2|v_{i}|+1)\right\rfloor +2\left\lfloor \log\left(\left\lfloor \log(2|v_{i}|+1)\right\rfloor +1\right)\right\rfloor \right)+n+1\\
 & \stackrel{\mathrm{(a)}}{\le}\log(|k|+1/2)+2\log\left(\log(|k|+1/2)+2\right)\\
 & \;\;\;\;+\sum_{i=1}^{n}\left(\log\left(\left|v_{i}+\frac{1}{2}\right|+\frac{1}{2}\right)+2\log\left(\log\left(\left|v_{i}+\frac{1}{2}\right|+\frac{1}{2}\right)+2\right)\right)+2n+2\\
 & \le\log(k+\tau+1/2)+2\log\left(\log(k+\tau+1/2)+2\right)\\
 & \;\;\;\;+\sum_{i=1}^{n}\left(\log\left(2^{k}\left(\left\Vert x\right\Vert _{\infty}+2^{-k}\right)\right)+2\log\left(\log\left(2^{k}\left(\left\Vert x\right\Vert _{\infty}+2^{-k}\right)\right)+2\right)\right)+2n+2\\
 & \le\log(k+\tau+1/2)+2\log\left(\log(k+\tau+1/2)+2\right)\\
 & \;\;\;\;+n\left(\log\left(2^{k}\left(\left\Vert x\right\Vert _{\infty}+\mathrm{V}_{n}^{1/n}(A)\right)\right)+2\log\left(\log\left(2^{k}\left(\left\Vert x\right\Vert _{\infty}+\mathrm{V}_{n}^{1/n}(A)\right)\right)+2\right)\right)+2n+2\\
 & =n\left(\log\left(\left\Vert x\right\Vert _{\infty}+\mathrm{V}_{n}^{1/n}(A)\right)+k+2+2\log\left(\log\left(\left\Vert x\right\Vert _{\infty}+\mathrm{V}_{n}^{1/n}(A)\right)+k+2\right)\right)\\
 & \;\;\;\;+\log(k+\tau+1/2)+2+2\log\left(\log(k+\tau+1/2)+2\right)\\
 & =n \ell_{\delta}\left(\log\left(\left\Vert x\right\Vert _{\infty}+\mathrm{V}_{n}^{1/n}(A)\right)+k+2\right)+\ell_{\delta}\left(\log(k+\tau+1/2)+2\right)
\end{aligned}
\]
where $(a)$ follows by the fact that $\left\lfloor \log(2|i|+1)\right\rfloor \le\log(\left|2i+1\right|+1)$.

Let $X\sim\mathrm{Unif}(A)$, and $V,K$ be such that $V\in D_{K}(A)$
and $X\in C_{K,V}$. Then we have
\begin{align*}
  \E\left[L(Enc(\mathrm{Unif}(A))\right]
 & =\E\left[L(g_{C}(V,K)\right]\\
 & \le n\E\left[\ell_{\delta}\left(\log\left(\left\Vert X\right\Vert _{\infty}+\mathrm{V}_{n}^{1/n}(A)\right)+K+2\right)\right]+\E\left[\ell_{\delta}\left(\log(K+\tau+1/2)+2\right)\right]\\
 & \le n \ell_{\delta}\left(\E\left[\log\left(\left\Vert X\right\Vert _{\infty}+\mathrm{V}_{n}^{1/n}(A)\right)\right]+\E[K]+2\right)+\ell_{\delta}\left(\log(\E[K]+\tau+1/2)+2\right)
\end{align*}
by Jensen's inequality and the concavity of $\ell_{\delta}$. We now proceed to
bound 
\begin{align*}
\E[K] & =\frac{1}{\mathrm{V}_{n}(A)}\sum_{k=-\infty}^{\infty}k\cdot2^{-nl}\left|D_{l}(A)\right|.
\end{align*}
Consider
\begin{align*}
\sum_{l=-\infty}^{k}2^{-nl}\left|D_{l}(A)\right| & =2^{-nk}\left|\left\{ v\in\mathbb{Z}^{n}:\,C_{k,v}\subseteq A\right\} \right|\\
 & \ge2^{-nk}\left|\left\{ v\in\mathbb{Z}^{n}:\,C_{k-1,\,(v-w)/2}\subseteq A\right\} \right|.
\end{align*}
for any $w\in[0,1]^{n}$, since $C_{k,v}\subseteq C_{k-1,\,(v-w)/2}$.
Note that the $(v-w)/2$ in the subscript may not have integer entries.
The same definition of $C_{k,v}=2^{-k}([0,1]^{n}+v)$ can still be
applied, however. Also
\begin{align*}
  \int_{[0,1]^{n}}\left|\left\{ v\in\mathbb{Z}^{n}:\,C_{k-1,\,(v-w)/2}\subseteq A\right\} \right|dw
 & =\sum_{v\in\mathbb{Z}^{n}}\int_{[0,1]^{n}}\mathbf{1}\left\{ C_{k-1,\,(v-w)/2}\subseteq A\right\} dw\\
 & =2^{n}\int_{\mathbb{R}^{n}}\mathbf{1}\left\{ C_{k-1,w}\subseteq A\right\} dw\\
 & =2^{n}2^{n(k-1)}\mathrm{V}_{n}\left(A\ominus[0,\,2^{-(k-1)}]^{n}\right)\\
 & =2^{nk}\mathrm{V}_{n}\left(A\ominus[0,\,2^{-(k-1)}]^{n}\right).
\end{align*}
Hence 
\[
\sum_{l=-\infty}^{k}2^{-nl}\left|D_{l}(A)\right|\ge\mathrm{V}_{n}\left(A\ominus[0,\,2^{-(k-1)}]^{n}\right),
\]
and
\[
\sum_{l=k+1}^{\infty}2^{-nl}\left|D_{l}(A)\right|\le\mathrm{V}_{n}(A)-\mathrm{V}_{n}\left(A\ominus[0,\,2^{-(k-1)}]^{n}\right).
\]
As a result,
\begin{align*}
\E[K] & =\frac{1}{\mathrm{V}_{n}(A)}\sum_{k=-\infty}^{\infty}k\cdot2^{-nl}\left|D_{l}(A)\right|\\
 & =\frac{1}{\mathrm{V}_{n}(A)}\sum_{k=-\infty}^{\infty}\left(\mathbf{1}\left\{ k\ge0\right\} \mathrm{V}_{n}(A)-\sum_{l=-\infty}^{k}2^{-nl}\left|D_{l}(A)\right|\right)\\
 & \le\frac{1}{\mathrm{V}_{n}(A)}\sum_{k=-\infty}^{\infty}\left(\mathbf{1}\left\{ k\ge0\right\} \mathrm{V}_{n}(A)-\mathrm{V}_{n}\left(A\ominus[0,\,2^{-(k-1)}]^{n}\right)\right)\\
 & \le\frac{1}{\mathrm{V}_{n}(A)}\int\left(\mathbf{1}\left\{ t\ge0\right\} \mathrm{V}_{n}(A)-\mathrm{V}_{n}\left(A\ominus[0,\,2^{-t}]^{n}\right)\right)dt+2\\
 & =h_{\ominus[0,1]^{n}}(A)+2.
\end{align*}
We have 
\[
\begin{aligned}\E\left[L(W)\right] & \le n \ell_{\delta}\left(h+\E\left[\log\left(\left\Vert X\right\Vert _{\infty}+\mathrm{V}_{n}^{1/n}(A)\right)\right]+4\right)+\ell_{\delta}\left(\log\left(h+2\max\left\{ \log\mathrm{V}_{n}^{1/n}(A),\,0\right\} +\frac{5}{2}\right)+2\right).\end{aligned}
\]
It remains to bound $\mathrm{V}_{n}^{1/n}(A)$ by $\E[\left\Vert X\right\Vert _{\infty}]$.
By the Markov inequality,
\begin{align}
\E[\left\Vert X\right\Vert _{\infty}] & \ge\frac{1}{4}\mathrm{V}_{n}^{1/n}(A)\cdot\mathrm{P}\left\{ \left\Vert X\right\Vert _{\infty}\ge\frac{1}{4}\mathrm{V}_{n}^{1/n}(A)\right\} \nonumber \\
 & =\frac{1}{4}\mathrm{V}_{n}^{1/n}(A)\cdot\frac{\mathrm{V}_{n}\left\{ x\in A:\,\left\Vert x\right\Vert _{\infty}\ge(1/4)\mathrm{V}_{n}^{1/n}(A)\right\} }{\mathrm{V}_{n}(A)}\nonumber \\
 & \ge\frac{1}{4}\mathrm{V}_{n}^{1/n}(A)\cdot\frac{\mathrm{V}_{n}(A)-\left((1/2)\mathrm{V}_{n}^{1/n}(A)\right)^{n}}{\mathrm{V}_{n}(A)}\nonumber \\
 & \ge\frac{1}{8}\mathrm{V}_{n}^{1/n}(A).\label{eq:norm_vol}
\end{align}
Hence,
\begin{align}
\E\left[L(W)\right] & \le n \ell_{\delta}\left(h+\log\E[\left\Vert X\right\Vert _{\infty}]+4+\log9\right)+\ell_{\delta}\left(\log\left(h+2\max\left\{ \log\E[\left\Vert X\right\Vert _{\infty}]+3,\,0\right\} +\frac{5}{2}\right)+2\right).\label{eq:thm_better}
\end{align}
This completes the proof of the theorem. 
\end{IEEEproof}
Combining Lemma \ref{lem:erosion_bd_ortho_norm} and Theorem \ref{thm:unif_len},
we can bound the expected length of the universal dyadic coding scheme for orthogonally convex sets. 
\begin{cor}
\label{cor:unif_len_ortho}
The expected codeword length of the universal dyadic coding scheme for uniform pdfs
applied to an orthogonally convex $A\subseteq\mathbb{R}^{n}$ is upper bounded as
\textnormal{
\[
\begin{aligned}\E\left[L(W)\right] & \le n \ell_{\delta}\left((n-1)\log r+\log(\left\Vert \hat{x}\right\Vert _{\infty}+r)-\log\mathrm{V}_{n}(A)+4n+8\right)\\
 & \;\;\;+\ell_{\delta}\left(\log\left((n-1)\log r+2\max\left\{ r,\,0\right\} -\log\mathrm{V}_{n}(A)+4n+9\right)+2\right).
\end{aligned}
\]
}
for any $\hat{x}\in\mathbb{R}^{n}$, where $\ell_{\delta}(t)=t+2\log t$ and \textnormal{$r=\E\left[\left\Vert X-\hat{x}\right\Vert _{\infty}\right]$}. 
\end{cor}

\begin{IEEEproof}[Proof of Corollary \ref{cor:unif_len_ortho}]
By Lemma \ref{lem:erosion_bd_ortho_norm}, 
\[
h_{\ominus[0,1]^{n}}(A)\le(n-1)\log\E\left[\left\Vert X-\hat{x}\right\Vert _{\infty}\right]-\log\mathrm{V}_{n}(A)+4n.
\]
By Theorem \ref{thm:unif_len}, 
\[
\begin{aligned}\E\left[L(W)\right] & \le n \ell_{\delta}\left(h+\log\left(\E[\left\Vert X\right\Vert _{\infty}]+\mathrm{V}_{n}^{1/n}(A)\right)+4\right)+\ell_{\delta}\left(\log\left(h+2\max\left\{ \log\mathrm{V}_{n}^{1/n}(A),\,0\right\} +\frac{5}{2}\right)+2\right)\\
 & \le n \ell_{\delta}\left(h+\log\E[\left\Vert X\right\Vert _{\infty}]+8\right)+\ell_{\delta}\left(\log\left(h+2\max\left\{ \log\E[\left\Vert X-\hat{x}\right\Vert _{\infty}],\,0\right\} +9\right)+2\right)\\
 & \le n \ell_{\delta}\left((n-1)\log\E\left[\left\Vert X-\hat{x}\right\Vert _{\infty}\right]+\log\E[\left\Vert X\right\Vert _{\infty}]-\log\mathrm{V}_{n}(A)+4n+8\right)\\
 & \;\;\;+\ell_{\delta}\left(\log\left((n-1)\log\E\left[\left\Vert X-\hat{x}\right\Vert _{\infty}\right]+2\max\left\{ \log\E[\left\Vert X-\hat{x}\right\Vert _{\infty}],\,0\right\} -\log\mathrm{V}_{n}(A)+4n+9\right)+2\right).
\end{aligned}
\]

\end{IEEEproof}

An added benefit of our universal dyadic coding scheme is that if $X$ can be generated in a distributed manner. Suppose $X$ is an $n$-dimensional vector $X_1,\ldots,X_n$ and 
instead of having one decoder wishing to generate $X$, we
have $n$ decoders that all receive $W$ and decoder $i$ wishes to generate only $X_{i}$, $i\in [1:n]$. Such distributed generation is possible using our universal dyadic coding scheme since decoder $i$ can
generate $X_{i}$ uniformly over the interval $[2^{-k}v_{i},\,2^{-k}(v_{i}+1)]$
without any need to cooperate with other decoders. In~\cite{distsimcont_arxiv}. we described a dyadic decomposition coding scheme for distributed generation of a given pdf. The scheme in this paper differs from that in~\cite{distsimcont_arxiv} in several aspects.
\begin{itemize}
\item The scheme in this paper is universal, while the scheme in~\cite{distsimcont_arxiv} is constructed for a given pdf known to both the encoder and the decoders.
\item In~\cite{distsimcont_arxiv} we used an optimal prefix free code, such as Huffman code, to encode the hypercubes, while in this paper we use a universal code over the integers since the distribution on the hypercubes is known a priori.
\item In~\cite{distsimcont_arxiv}, we can perform scaling (and bijective transformations) on each variable $X_i$ before applying the dyadic decomposition scheme. It is not possible to perform such preprocessing here since the decoder would not know the scaling factor or the bijective transformation used.
\item In the analysis of the expected codeword length
%(or equivalently, the entropy of the dyadic decomposition, since the distribution is fixed)
in~\cite{distsimcont_arxiv}, it suffices to consider only the distribution of the sizes of the hypercubes. In our universal scheme, both the size and the position of the hypercube affect the length of the codeword assigned to it. %This makes the analysis different from that in~\cite{distsimcont_arxiv}
%\item In the analysis of ???? in~\cite{distsimcont_arxiv}, we only need to consider the entropy of the dyadic decomposition, since it suffices to consider only the distribution of the sizes of the hypercubes. In our universal scheme, both the size and the position of the hypercube affect the length of the codeword assigned to it. %This makes the analysis different from that in~\cite{distsimcont_arxiv}
\end{itemize}

\section{Non-uniform Distributions\label{sec:nonunif}}

In this section, we extend the results of the previous section to the case where the pdf of $X=(X_{1},\ldots,X_{n})$
is selected from a set of arbitrary (not necessarily uniform) pdfs. The key idea in extending our scheme is the following.
Note that in general, any pdf can be written as a mixture of uniform pdfs.
Let $Z\sim f_{Z}$, where $f_{Z}(z)=\mathrm{V}_{n}(L_{z}^{+}(f))$
for $z\ge0$ and $L_{z}^{+}(f)=\left\{ x\in\mathbb{R}^{n}:\,f(x)\ge z\right\} $
is the superlevel set of $f$. Let $X|\{Z=z\} \sim \U(L_{z}^{+}(f))$, then we have $X \sim f(x)$. Hence $f(x)$ can be expressed as a mixture of uniform distributions over $L_{z}^{+}(f)$ for different values of $z$. Alice can first generate $Z\sim f_{Z}$, then apply the universal dyadic coding scheme for uniform
distributions on $L_{Z}^{+}(f)$. The scheme is formally defined as follows. 
 
%universal dyadic coding scheme on the set $A_{1}$ or $A_{2}$
%at random with probabilities 1/2, then the output distribution will
%be the mixture of the uniform distributions $\U(A_{1})$
%and $\U(A_{2})$. Using this idea, we can construct
%any pdf using a mixture of uniform distributions.
%Let $Z\sim f_{Z}$, where $f_{Z}(z)=\mathrm{V}_{n}(L_{z}^{+}(f))$
%for $z\ge0$ and $L_{z}^{+}(f)=\left\{ x\in\mathbb{R}^{n}:\,f(x)\ge z\right\} $
%is the superlevel set of $f$. We generate $Z\sim f_{Z}$,
%and then apply the universal dyadic coding scheme for uniform
%distributions on $L_{Z}^{+}(f)$. It is straightforward to check that
%$(X_{1},\ldots,X_{n})\sim f$. The scheme is formally defined as follows. 

\noindent {\bf Universal dyadic coding scheme for general pdfs}.
The universal dyadic coding scheme for the set of almost everywhere continuous pdfs $\Pr$ consists of:
\begin{enumerate}
\item A stochastic encoder that generates $\xt$ according to the observed $f$ and $Z\sim \U[0,f(\xt)]$, and
finds $(k,v)$ such that $v\in D_{k}(L_{z}^{+}(f))$ and $\xt\in C_{k,v}$. The encoder maps $(k,v)$ into a codeword $w$ that consists of the concatenation of the signed Elias delta codewords for $k,v_{1},\ldots,v_{n}$, i.e., 
$w=g_{C}(k,v)=g_{\delta}(k)\,\Vert\, g_{\delta}(v_{1})\,\Vert\,\cdots\,\Vert\, g_{\delta}(v_{n})$.
\item A stochastic decoder that upon receiving $w$ recovers $(v,k)$ and generates $x$ uniformly over $C_{k,v}$.
\end{enumerate}

We illustrate this scheme in the following.
\begin{example}
Assume that the selected pdf is Gaussian with zero mean and unit variance. 
Figure~\ref{fig:nonunif_gauss} depicts the universal dyadic coding scheme for this pdf. The horizontal and vertical axes represent $x$ and $z$, respectively. The encoder sends the codeword for the rectangle containing $(x,z)$. The expected codeword length (computed by listing all intervals in the dyadic decomposition with length at least $2^{-20}$) is $7.06$.
\begin{figure}[h!]
\begin{center}
\includegraphics[scale=0.28]{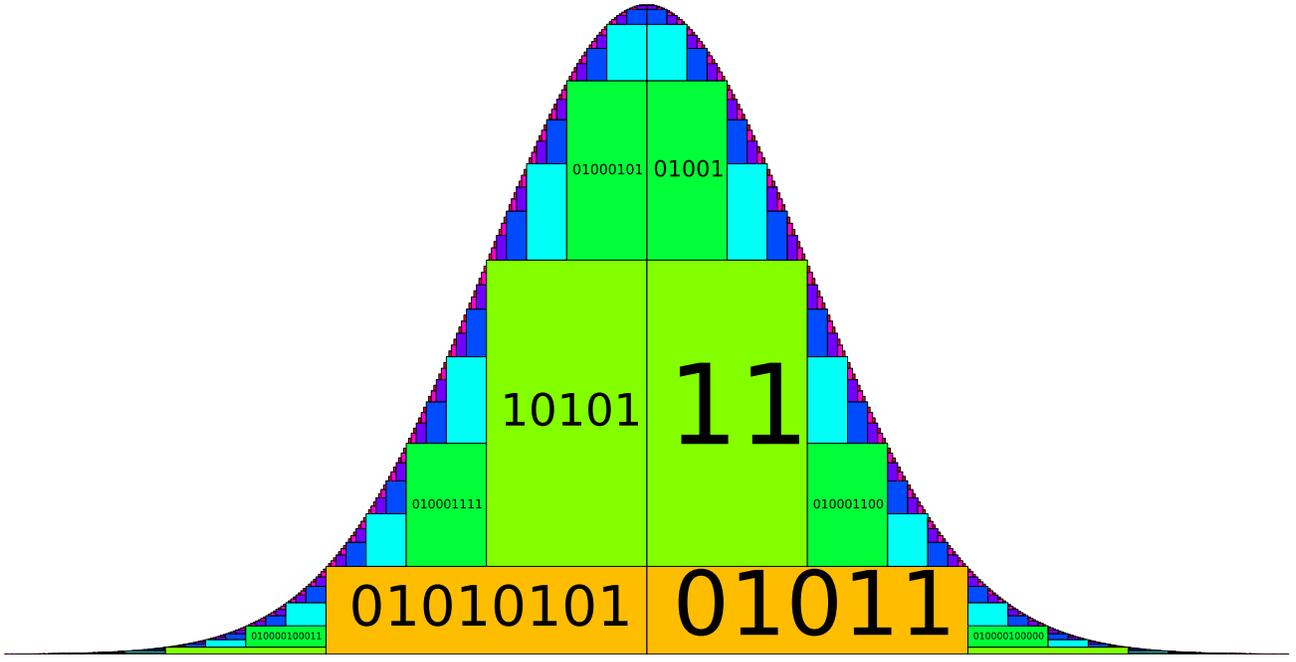}
\caption{Universal dyadic coding scheme on $\mathcal{N}(0,1)$.}
\label{fig:nonunif_gauss} 
\end{center}
\end{figure}
\end{example}

As a consequence of Theorem~\ref{thm:unif_len}, we have the following
bound on the expected codeword length.
\begin{thm}
\label{thm:nonunif_len}The expected codeword length of the universal dyadic coding scheme for $X\sim f(x)$ is upper bounded as
\textnormal{
\[
\begin{aligned}\E\left[L(W)\right] & \le n \ell_{\delta}\left(\E_{Z}\left[h\right]+\log\E[\left\Vert X\right\Vert _{\infty}]+8\right)+\ell_{\delta}\left(\log\left(\E_{Z}\left[h\right]+2\max\left\{ \log\E[\left\Vert X\right\Vert _{\infty}],\,0\right\} +10\right)+2\right),\end{aligned}
\]
}
where $\ell_{\delta}(t)=t+2\log t$ and $h=h_{\ominus[0,1]^{n}}(L_{Z}^{+}(f))$
is a random variable, where $Z\sim f_{Z}$, $f_{Z}(z)=\mathrm{V}_{n}(L_{z}^{+}(f))$
for $z\ge0$.\end{thm}
\begin{IEEEproof}
%\textcolor{blue}{First we check that $L_{z}^{+}(f)$ has a boundary of measure zero for almost all $z$, in order for the coding scheme to succeed almost surely. By the coarea formula for functions of bounded variation~\cite{miranda2003functions},
%\[
%\int_0^{\infty} \Vert \partial L_{z}^{+}(f) \Vert dt < \infty,
%\]
%where $\Vert \partial L_{z}^{+}(f) \Vert$ is the perimeter of $L_{z}^{+}(f)$ as a Caccioppoli set. Hence $\Vert \partial L_{z}^{+}(f) \Vert < \infty$ for almost all $z$, which implies $\mathrm{V}_n(\partial L_{z}^{+}(f))< \infty$ for almost all $z$.}
Using~\eqref{eq:thm_better}, 
\[
\begin{aligned}\E\left[L(W)\right] & \le n\E_{Z}\left[\ell_{\delta}\left(h+\log\E[\left\Vert X\right\Vert _{\infty}|\,Z]+8\right)\right]+\E_{Z}\left[\ell_{\delta}\left(\log\left(h+2\max\left\{ \log\E[\left\Vert X\right\Vert _{\infty}],\,0\right\} +8.5\right)+2\right)\right]\\
 & \le n \ell_{\delta}\left(\E_{Z}\left[h\right]+\log\E[\left\Vert X\right\Vert _{\infty}]+8\right)+\ell_{\delta}\left(\log\left(\E_{Z}\left[h\right]+2\E_{Z}\left[\max\left\{ \log\E[\left\Vert X\right\Vert _{\infty}|\,Z],\,0\right\} \right]+8.5\right)+2\right).
\end{aligned}
\]
To bound $\E_{Z}\left[\max\left\{ \log\E[\left\Vert X\right\Vert _{\infty}|\,Z],\,0\right\} \right]$,
define a concave function $q:[0,\infty)\to[0,\infty)$, 
\[
q(t)=\begin{cases}
te^{-1}\log e & \text{if}\;t\le e\\
\log t & \text{if}\;t>e.
\end{cases}
\]
Note that 
\[
\max\{\log t,\,0\}\le q(t)\le\max\{\log t,\,0\}+e^{-1}\log e.
\]
Hence,
\begin{align*}
  \E_{Z}\left[\max\left\{ \log\E[\left\Vert X\right\Vert _{\infty}|\,Z],\,0\right\} \right]
 & \le\E_{Z}\left[q(\E[\left\Vert X\right\Vert _{\infty}|\,Z])\right]\\
 & \le q(\E[\left\Vert X\right\Vert _{\infty}])\\
 & \le\max\left\{ \log\E[\left\Vert X\right\Vert _{\infty}],\,0\right\} +e^{-1}\log e.
\end{align*}
Note that we also need $L_{z}^{+}(f)$ to have a boundary of measure zero for almost all $z$, in order for the coding scheme to succeed almost surely. This is implied by the almost everywhere continuity of $f(x)$. The proof of this claim is given in Appendix~\ref{apx:nonunif_len}.
\end{IEEEproof}
We can also generalize Corollary \ref{cor:unif_len_ortho} to orthogonally
concave pdfs (which includes quasiconcave pdfs) as follows. 
\begin{cor}
\label{cor:nonunif_len_ortho}The expected codeword length of the universal dyadic coding scheme
for $X\sim f(x)$, where $f$ is orthogonally concave, is upper bounded as
\textnormal{
\[
\begin{aligned}\E\left[L(W)\right] & \le n \ell_{\delta}\left((n-1)\log r+\log(\left\Vert \hat{x}\right\Vert _{\infty}+r)+h(Z)+4n+8\right)\\
 & \;\;\;+\ell_{\delta}\left(\log\left((n-1)\log r+2\max\left\{\log r,\,0\right\} +h(Z)+4n+10\right)+2\right).
\end{aligned}
\]
}
for any $\hat{x}\in\mathbb{R}^{n}$, where $\ell_{\delta}(t)=t+2\log t$, \textnormal{$r=\E\left[\left\Vert X-\hat{x}\right\Vert _{\infty}\right]$},
$Z\sim f_{Z}$, $f_{Z}(z)=\mathrm{V}_{n}(L_{z}^{+}(f))$ for $z\ge0$.
As a result, if $\sup_{x}f(x)<\infty$, 
\textnormal{
\[
\begin{aligned}\E\left[L(W)\right] & \le n \ell_{\delta}\left((n-1)\log r+\log(\left\Vert \hat{x}\right\Vert _{\infty}+r)+\log\sup_{x}f(x)+4n+8\right)\\
 & \;\;\;+\ell_{\delta}\left(\log\left((n-1)\log r+2\max\left\{\log r,\,0\right\} +\log\sup_{x}f(x)+4n+10\right)+2\right).
\end{aligned}
\]
}
\end{cor}

\section{Bounded Support Distributions\label{sec:bounded}}

In this section, we present a variant of the universal dyadic coding scheme for a set of distributions with a uniform bound on their support.
Without loss of generality, assume $\Pr$
consists of the set of pdfs over $[0,1]^{n}$.
Since the $v$ in the definition of our universal dyadic coding scheme (corresponding to the position of the hypercube) is bounded, we can use a fixed length code to encode $(v_1,\ldots,v_n)$. This allows us to reduce the expected codeword length.

\noindent {\bf Universal dyadic coding scheme for pdfs over the unit hypercube}.
The universal dyadic coding scheme for pdfs over $[0,1]^{n}$ consists of: 
\begin{enumerate}
\item A stochastic encoder that generates $\tilde{x}$ according to the observed $f$ and $z\sim\U[0,f(\tilde{x})]$ and
finds $(k,v)$ such that $v\in D_{k}(L_{z}^{+}(f))$ and $\tilde{x}\in C_{k,v}$.  The encoder then maps $(k,v)$ into a codeword which consists of the concatenation
of the unsigned Elias gamma codeword for $k+1$, and the $k$-bit binary
representations of $v_{1},\ldots,v_{n}$, i.e., 
$w=g_{C}(k,v)=g_{\gamma+}(k+1)\,\Vert\, g_{\mathrm{b},k}(v_{1})\,\Vert\,\cdots\,\Vert\, g_{\mathrm{b},k}(v_{1})$,
where $g_{\mathrm{b},k}(i)$ is the binary representation of $i$
with $k$ bits, possibly with leading zeros.
\item A stochastic decoder that upon observing $w$ recovers $(v,k)$ and generates $x$
uniformly over $C_{k,v}$. 
\end{enumerate}

\begin{figure}[h!]
\begin{center}
\includegraphics[scale=0.45]{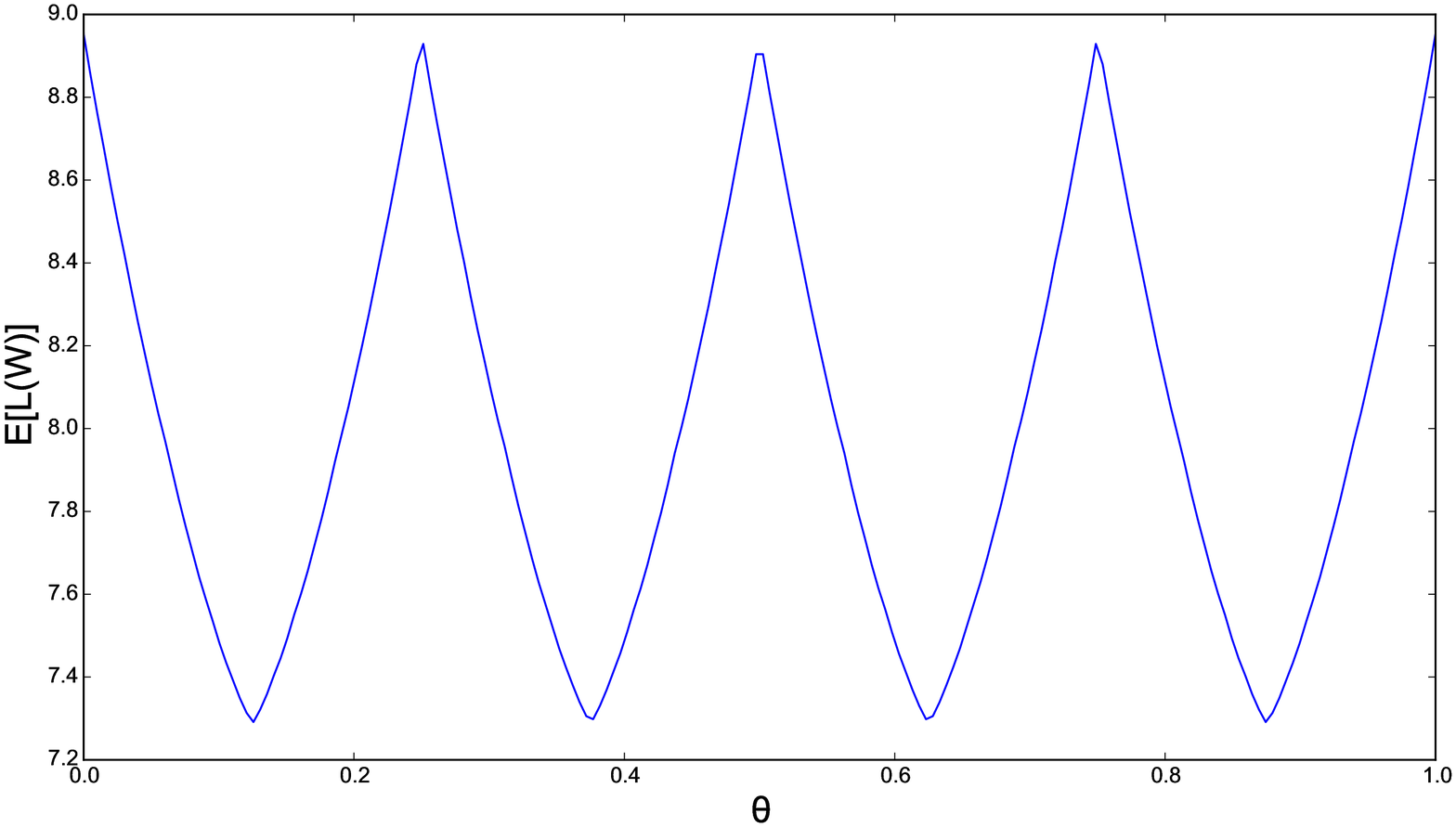}
\caption{Universal dyadic coding scheme for pdfs over the unit hypercube applied to the simulation of the Bell state.}
\label{fig:poscos} 
\end{center}
\end{figure}

Since the length of the unsigned Elias gamma codeword $g_{\gamma}(k+1)$
is $2\left\lfloor \log(k+1)\right\rfloor +1$, the length of $w$ is
\[
L(w)=nk+2\left\lfloor \log(k+1)\right\rfloor +1.
\]
The expected codeword length is upper bounded as follows. 
\begin{thm}
\label{thm:bdd_len}
The expected codeword length of the universal dyadic coding scheme for pdfs over the unit hypercube
for $X\sim f(x)$, where $f$ is orthogonally concave, is upper bounded as
\textnormal{
\[
\begin{aligned}\E\left[L(W)\right] & \le n\left(h(Z)+\log n+\log e+2\right)+2\log\left(h(Z)+\log n+\log e+3\right)+1,\end{aligned}
\]
}
 where $Z\sim f_{Z}$, $f_{Z}(z)=\mathrm{V}_{n}(L_{z}^{+}(f))$ for
$z\ge0$. As a result, if $\sup_{x}f(x)<\infty$,
\textnormal{
\[
\begin{aligned}\E\left[L(W)\right] & \le n\left(\log\sup_{x}f(x)+\log n+\log e+2\right)+2\log\left(\log\sup_{x}f(x)+\log n+\log e+3\right)+1.\end{aligned}
\]
}
\end{thm}
\begin{IEEEproof}
In~\cite{distsimcont_arxiv} (Theorem 1), it was shown that the erosion entropy for orthogonally convex $A$ is bounded as
\[
h_{\ominus[0,1]^{n}}(A)\le\log\left(\frac{\sum_{i=1}^{n}\mathrm{V}_{n-1}\mathrm{P}_{\backslash i}(A)}{\mathrm{V}_{n}(A)}\right)+\log e,
\]
where $\mathrm{VP}_{\backslash i}(A)=\left\{ (x_{1},\ldots,x_{i-1},x_{i+1},\ldots,x_{n}):\, x\in A\right\} $.
Let $Z\sim f_{Z}$, $f_{Z}(z)=\mathrm{V}_{n}(L_{z}^{+}(f))$, then
\[
h_{\ominus[0,1]^{n}}(L_{z}^{+}(f))\le-\log\mathrm{V}_{n}(L_{z}^{+}(f))+\log n+\log e.
\]
Hence
\[
\begin{aligned}\E\left[L(W)\right] & \le n\E[K]+2\log(\E[K]+1)+1\\
 & \le n\left(\E[h_{\ominus[0,1]^{n}}(L_{Z}^{+}(f))]+2\right)+2\log\left(\E[h_{\ominus[0,1]^{n}}(L_{Z}^{+}(f))]+3\right)+1\\
 & \le n\left(\E[-\log\mathrm{V}_{n}(L_{Z}^{+}(f))]+\log n+\log e+2\right)+2\log\left(\E[-\log\mathrm{V}_{n}(L_{Z}^{+}(f))]+\log n+\log e+3\right)+1\\
 & =n\left(h(Z)+\log n+\log e+2\right)+2\log\left(h(Z)+\log n+\log e+3\right)+1.
\end{aligned}
\]

\end{IEEEproof}
As an example, we apply this result to 
simulating the Bell state in Application~\ref{ex:simquantum}
\[
f(x\,|\,\theta)=\pi\max\{\cos(2\pi(x-\theta)),\,0\},
\]
fitted to the interval $[0,1]$. Although this pdf is not orthogonally
concave, it can be decomposed into at most two
orthogonally concave parts with disjoint support, hence the expected
codeword length is the weighted average of the expected codeword lengths
for those two pieces, which incurs a penalty of at most 1 bit. By Theorem~\eqref{thm:bdd_len},
\[
\begin{aligned}\E\left[L(W)\right] & \le\log\pi+\log e+2+2\log\left(\log\pi+\log e+3\right)+2\approx12.31.\end{aligned}
\]
Figure~\ref{fig:poscos} plots the numerical values of $\E\left[L(W)\right]$ versus $\theta$ computed by listing all intervals in the dyadic decomposition with length at least $2^{-17}$. As can be seen, $\E\left[L(W)\right] \le 8.96$ for all $\theta$.

\section{Lower Bound on Expected Codeword Length\label{sec:lbound}}

In the previous sections we focused on schemes for universal remote generation of continuous random variables and upper bounds on their expected codeword length.  In this section, we establish a lower bound on the expected codeword
length that every remote generation scheme must satisfy. 

Consider a universal remote generation coding scheme with a prefix-free codeword
set $C\subseteq\{0,1\}^{*}$. Upon receiving $w\in C$, Bob generates
$X|\{W=w\}\sim\tilde{p}_{w}$ according to a distribution $\tilde{p}_{w}(dx)$. Define the \emph{implied distribution}
of this scheme as
\[
p_{\mathrm{Im}}(dx)=\Big(\sum_{w'\in C}2^{-L(w')}\Big)^{-1}\sum_{w\in C}2^{-L(w)}\tilde{p}_{w}(dx).
\]
We now show that the expected codeword length $\E_{p}(L(W))$ is lower bounded by the relative entropy between $p$
and $p_{\mathrm{Im}}$.
\begin{thm}
\label{thm:lbound}For a universal remote generation scheme with an implied distribution $p_{\mathrm{Im}}$, the average codeword length for $p\in \Pr$ is lower bounded as
\textnormal{
\[
\E_{p}(L(W))\ge D(p\Vert p_{\mathrm{Im}}).
\]
}
\end{thm}
\begin{IEEEproof}
Consider the input distribution $p$. Assume the encoder outputs $w$
with probability $a(w)$. Then $p=\sum_{w\in C}a(w)\tilde{p}_{w}$,
$\E_{p}(L(W))=\sum_{w\in C}a(w)L(w)$. By convexity of relative entropy,
\[
\begin{aligned}D(p\Vert p_{\mathrm{Im}}) & =D\left(\sum_{w\in C}a(w)\tilde{p}_{w}\,\Vert\,p_{\mathrm{Im}}\right)\\
 & \le\sum_{w\in C}a(w)D\left(\tilde{p}_{w}\Vert p_{\mathrm{Im}}\right)\\
 & \le\sum_{w\in C}a(w)L(w)\\
 & =\E_{p}(L(W)).
\end{aligned}
\]

\end{IEEEproof}

Consider the (unbounded support) dyadic universal code. The implied distribution in this case is described by a pdf
$f_{\mathrm{Im}}(x)\propto\left\Vert x\right\Vert _{\infty}^{-n}\left(\log\left\Vert x\right\Vert _{\infty}\right)^{-2n}$.
Theorem \ref{thm:lbound} gives the lower bound on the expected codeword length for generating $X \sim f$,
\[
D\left(f_{X}\,\Vert\,f_{\mathrm{Im}}\right)\approx h(X)+n\E\left[\ell_{\delta}(\log\left\Vert X\right\Vert _{\infty})\right].
\]
Comparing this to Theorem \ref{thm:unif_len}, we see that the upper
bound is close to the lower bound when $\left\Vert X\right\Vert _{\infty}$
is the dominant term.

Note that Theorem~\ref{thm:lbound} continues to hold even when Alice and Bob are allowed to share unlimited common randomness (denoted by the random variable $Q$). %The implied distribution in this case is defined as follows.
Suppose the prefix-free codeword
set when $Q=q$ is $C_q\subseteq\{0,1\}^{*}$. Upon receiving $w\in C$, Bob generates
$X|\{Q=q,W=w\}\sim\tilde{p}_{q,w}$. The implied distribution is
\[
p_{\mathrm{Im}}(dx)=\int_{\mathcal{Q}} \Big(\sum_{w'\in C_q}2^{-L(w')}\Big)^{-1}\sum_{w\in C_q}2^{-L(w)}\tilde{p}_{q,w}(dx) p_Q(dq).
\]
Theorem~\ref{thm:lbound} still holds due to the convexity of relative entropy. Comparing this lower bound to the average length of the rejection sampling scheme in~\cite{harsha2010communication} (which requires common randomness), which achieves
\[
\E_{p}(L(W))\le D(p\Vert p^{*}) + 2 \log ( D(p\Vert p^{*})+1)+O(1)
\]
for some $p^{*}$. Hence, the lower bound is quite tight when unlimited common randomness is allowed.

\appendix
%dummy comment inserted by tex2lyx to ensure that this paragraph is not empty

\subsection{Proof of Lemma \ref{lem:erosion_bd_ortho_norm}} \label{apx:erosion_bd_ortho_norm}

The lemma is trivial when $n=1$ since $A$ can only be an interval.
Hence we assume $n\ge2$.

From the definition of erosion entropy, {\allowdisplaybreaks
\begin{align*}
h_{\ominus[0,1]^{n}}(A) & =\int_{-\infty}^{\infty}\left(\mathbf{1}\left\{ t\ge0\right\} -\frac{\mathrm{V}_{n}\left(A\ominus2^{-t}[0,1]^{n}\right)}{\mathrm{V}_{n}(A)}\right)dt\\
 & =\int_{-\infty}^{\infty}t\cdot d\left(\frac{\mathrm{V}_{n}\left(A\ominus2^{-t}[0,1]^{n}\right)}{\mathrm{V}_{n}(A)}\right)\\
 & \stackrel{\mathrm{(i)}}{=}\frac{1}{\mathrm{V}_{n}(A)}\int_{0}^{\infty}\left(\log s\right)\cdot d\mathrm{V}_{n}\left(A\ominus[0,s]^{n}\right)\\
 & =\frac{1}{\mathrm{V}_{n}(A)}\int_{0}^{\infty}\left(\log s\right)\cdot d\left(\sum_{i=1}^{n}\left(\mathrm{V}_{n}\left(A\ominus\{0\}^{n-i}\times[0,s]^{i}\right)-\mathrm{V}_{n}\left(A\ominus\{0\}^{n-i+1}\times[0,s]^{i-1}\right)\right)\right)\\
 & =\frac{1}{\mathrm{V}_{n}(A)}\int_{0}^{\infty}\left(-\log s\right)\left(\sum_{i=1}^{n}\mathrm{V}_{n-1}\mathrm{P}_{\backslash i}\left(A\ominus[0,s]^{n}\right)\right)ds
\end{align*}}
where (i) is by substituting $s=2^{-t}$, and $\mathrm{P}_{\backslash i}(A)=\left\{ (x_{1},\ldots,x_{i-1},x_{i+1},\ldots,x_{n})\,:\,x\in A\right\} \subseteq\mathbb{R}^{n-1}$.
Let 
\[
q(s)=\frac{\sum_{i=1}^{n}\mathrm{V}_{n-1}\mathrm{P}_{\backslash i}\left(A\ominus[0,s]^{n}\right)}{\mathrm{V}_{n}(A)}.
\]
Then we have $\int_{0}^{\infty}q(s)ds=1$, $\int_{0}^{\infty}(-\log s)q(s)ds=h_{\ominus[0,1]^{n}}(A)$.
Also {\allowdisplaybreaks
\begin{align*}
\E\left[\left\Vert X\right\Vert _{\infty}\right] & =\frac{1}{\mathrm{V}_{n}(A)}\int_{A}\left\Vert x\right\Vert _{\infty}dx\\
 & =\frac{-1}{\mathrm{V}_{n}(A)}\int_{0}^{\infty}\frac{d}{ds}\left(\int_{A\ominus[0,s]^{n}}\left\Vert x\right\Vert _{\infty}dx\right)ds\\
 & =\frac{-1}{\mathrm{V}_{n}(A)}\int_{0}^{\infty}\sum_{i=1}^{n}\left.\frac{\partial}{\partial s_{i}}\left(\int_{A\ominus[0,s_{1}]\times\cdots[0,s_{n}]}\left\Vert x\right\Vert _{\infty}dx\right)\right|_{(s_{1},\ldots,s_{n})=(s,\ldots,s)}ds\\
 & \ge\frac{-1}{\mathrm{V}_{n}(A)}\int_{0}^{\infty}\sum_{i=1}^{n}\left.\frac{\partial}{\partial s_{i}}\left(\int_{A\ominus[0,s_{1}]\times\cdots[0,s_{n}]}\left\Vert x_{[1:n]\backslash i}\right\Vert _{\infty}dx\right)\right|_{(s_{1},\ldots,s_{n})=(s,\ldots,s)}ds\\
 & =\frac{1}{\mathrm{V}_{n}(A)}\int_{0}^{\infty}\left(\sum_{i=1}^{n}\int_{\mathrm{P}_{\backslash i}\left(A\ominus[0,s]^{n}\right)}\left\Vert x_{[1:n]\backslash i}\right\Vert _{\infty}dx_{[1:n]\backslash i}\right)ds\\
 & \stackrel{\mathrm{(i)}}{\ge}\frac{1}{\mathrm{V}_{n}(A)}\int_{0}^{\infty}\left(\sum_{i=1}^{n}\frac{1}{8}\mathrm{V}_{n-1}\mathrm{P}_{\backslash i}^{1+1/(n-1)}\left(A\ominus[0,s]^{n}\right)\right)ds\\
 & \ge\frac{1}{8\mathrm{V}_{n}(A)}\int_{0}^{\infty}n^{-1/(n-1)}\left(\sum_{i=1}^{n}\mathrm{V}_{n-1}\mathrm{P}_{\backslash i}\left(A\ominus[0,s]^{n}\right)\right)^{n/(n-1)}ds\\
 & =\frac{1}{8}n^{-1/(n-1)}\mathrm{V}_{n}^{1/(n-1)}(A)\int_{0}^{\infty}q^{n/(n-1)}(s)ds,
\end{align*}
where (i) is by (\ref{eq:norm_vol}) in the proof of Theorem~(\ref{thm:unif_len}).
Let
\[
\tilde{q}(s)=\begin{cases}
\left(\beta(\log e)^{n-1}\Gamma(n)\right)^{-1}\left(-\log\left(s/\beta\right)\right)^{n-1} & \text{if}\;s\le\beta\\
0 & \text{if}\;s>\beta,
\end{cases}
\]
where $\beta=e^{n}2^{-h}$, $h=h_{\ominus[0,1]^{n}}(A)$. Then $\int_{0}^{\infty}\tilde{q}(s)ds=1$,
$\int_{0}^{\infty}(-\log s)\tilde{q}(s)ds=h$. Hence,
\begin{align*}
  \int_{0}^{\infty}q^{n/(n-1)}(s)ds
 & \ge\int_{0}^{\beta}q^{n/(n-1)}(s)ds\\
 & =\int_{0}^{\beta}\tilde{q}(s)^{n/(n-1)}\left(q(s)/\tilde{q}(s)\right)^{n/(n-1)}ds\\
 & \stackrel{\mathrm{(i)}}{\ge}\left(\int_{0}^{\beta}\tilde{q}(s)^{n/(n-1)}ds\right)\left(\left(\int_{0}^{\beta}\tilde{q}(s)^{n/(n-1)}ds\right)^{-1}\int_{0}^{\beta}\tilde{q}(s)^{n/(n-1)}\left(q(s)/\tilde{q}(s)\right)ds\right)^{n/(n-1)}\\
 & =\left(\int_{0}^{\beta}\tilde{q}(s)^{n/(n-1)}ds\right)^{-1/(n-1)}\left(\int_{0}^{\beta}\tilde{q}(s)^{1/(n-1)}q(s)ds\right)^{n/(n-1)}\\
 & =\left(n\left(\beta\Gamma(n)\right)^{-1/(n-1)}\right)^{-1/(n-1)}\left(\left(\log e\right)^{-1}\left(\beta\Gamma(n)\right)^{-1/(n-1)}\int_{0}^{\beta}\left(-\log s+\log\beta\right)q(s)ds\right)^{n/(n-1)}\\
 & \ge\left(n\left(\beta\Gamma(n)\right)^{-1/(n-1)}\right)^{-1/(n-1)}\left(\left(\log e\right)^{-1}\left(\beta\Gamma(n)\right)^{-1/(n-1)}\int_{0}^{\infty}\left(-\log s+\log\beta\right)q(s)ds\right)^{n/(n-1)}\\
 & =\left(n\left(\beta\Gamma(n)\right)^{-1/(n-1)}\right)^{-1/(n-1)}\left(\left(\log e\right)^{-1}\left(\beta\Gamma(n)\right)^{-1/(n-1)}\left(h+\log\beta\right)\right)^{n/(n-1)}\\
 & =n\left(\beta\Gamma(n)\right)^{-1/(n-1)},
\end{align*}
where (i) is by weighted power mean inequality. As a result,
\begin{align*}
\E\left[\left\Vert X\right\Vert _{\infty}\right] & \ge\frac{1}{8}n^{-1/(n-1)}\mathrm{V}_{n}^{1/(n-1)}(A)\int_{0}^{\infty}q^{n/(n-1)}(s)ds\\
 & \ge\frac{1}{8}n^{(n-2)/(n-1)}\mathrm{V}_{n}^{1/(n-1)}(A)\left(e^{n}2^{-h}\Gamma(n)\right)^{-1/(n-1)},
\end{align*}
\begin{align*}
h & \le(n-1)\left(\log\E\left[\left\Vert X\right\Vert _{\infty}\right]-\log\left(\frac{1}{8}n^{(n-2)/(n-1)}\mathrm{V}_{n}^{1/(n-1)}(A)\right)\right)+\log\Gamma(n)+n\log e\\
 & =(n-1)\log\E\left[\left\Vert X\right\Vert _{\infty}\right]-\log\mathrm{V}_{n}(A)+3(n-1)-(n-2)\log n+\log\Gamma(n)+n\log e\\
 & \le(n-1)\log\E\left[\left\Vert X\right\Vert _{\infty}\right]-\log\mathrm{V}_{n}(A)+3(n-1)-(n-2)\log n+\left(n\log n-(n-1)\log e\right)+n\log e\\
 & =(n-1)\log\E\left[\left\Vert X\right\Vert _{\infty}\right]-\log\mathrm{V}_{n}(A)+3(n-1)+2\log n+\log e\\
 & \le(n-1)\log\E\left[\left\Vert X\right\Vert _{\infty}\right]-\log\mathrm{V}_{n}(A)+4n.
\end{align*}}

\subsection{Proof of the claim on measure zero boundary in Theorem \ref{thm:nonunif_len}} \label{apx:nonunif_len}

We will prove that if $f$ is a pdf which is continuous almost everywhere, then $L_{z}^{+}(f)$ has a boundary of measure zero for almost all $z$. Assume the contrary that there exist an uncountable $G \subseteq [0,\infty)$ such that $\mathrm{V}_n (\partial L_{z}^{+}(f)) > 0$ for all $z \in G$ (note that $\partial L_{z}^{+}(f)$ is a Borel set and thus measurable). Then we show that there exists $z_1\neq z_2 \in G$ such that $\mathrm{V}_n (\partial L_{z_1}^{+}(f) \cap \partial L_{z_2}^{+}(f)) > 0$ (which follows from $\sigma$-finiteness, though we include a proof here for completeness). To show the claim, note that for any $z \in G$, there exists a hypercube $[0,1]^n + v$, $v \in \mathbb{Z}^n$ such that $\partial L_{z}^{+}(f) \cap ([0,1]^n + v)$ has nonzero measure. Hence there exists a hypercube $[0,1]^n + v$ such that $\partial L_{z}^{+}(f) \cap ([0,1]^n + v)$ has nonzero measure for an uncountable set of $z$'s. Since an uncountable collection of positive numbers must contain a finite subcollection with sum greater than 1, we can select $z_1,\ldots,z_m$ such that $\sum_i \mathrm{V}_n (\partial L_{z_i}^{+}(f) \cap ([0,1]^n + v)) > 1$, and hence there exists two of these sets with an intersection of nonzero measure.

Now we have $z_1 < z_2 \in G$ such that $\mathrm{V}_n (\partial L_{z_1}^{+}(f) \cap \partial L_{z_2}^{+}(f)) > 0$. Assume there exists $x$ in the intersection at which $f$ is continuous, since $x \in \partial L_{z_2}^{+}(f)$, there exist sequence $y_i \to x$ with $f(y_i) \ge z_2$, and hence $f(x) \ge z_2$. Also since $x \in \partial L_{z_1}^{+}(f) \subseteq \mathrm{cl}\{y : \, f(y) < z_1\}$, there exist sequence $\tilde{y}_i \to x$ with $f(\tilde{y}_i) <z_1$, and hence $f(x) \le z_1$, leading to a contradiction. Therefore $f$ is discontinuous in $\partial L_{z_1}^{+}(f) \cap \partial L_{z_2}^{+}(f)$, contradicting the assumption that $f$ is continuous almost everywhere. Therefore $L_{z}^{+}(f)$ has a boundary of measure zero for almost all $z$.

 \bibliographystyle{IEEEtran}
\bibliography{ref,nit}

\end{document}

%% file: defns.tex
\usepackage{xspace}
\usepackage{bbm}
\input{mathlig}

\newcommand{\muspace}{\mspace{1mu}}

\DeclareRobustCommand{\scond}{\mathchoice{\muspace\vert\muspace}{\vert}{\vert}{\vert}}
\mathlig{|}{\scond}

\DeclareRobustCommand{\discint}{\mathchoice{\mspace{-1.5mu}:\mspace{-1.5mu}}{\mspace{-1.5mu}:\mspace{-1.5mu}}{:}{:}}
\mathlig{::}{\discint}

\def\sgn{\mathop{\rm sgn}\nolimits}%
\renewcommand{\Pr}{\mathscr{P}}

\newcommand{\xt}{{\tilde{x}}}

\DeclareMathOperator\E{\textsf{E}}
\let\P\relax
\DeclareMathOperator\P{\textsf{P}}
\newcommand{\U}{\mathrm{Unif}}

\def\textiid{i.i.d.\@\xspace}
\newcommand\iid{\ifmmode\text{ i.i.d. } \else \textiid \fi}

\def\mathllap{\mathpalette\mathllapinternal}
\def\mathllapinternal#1#2{%
  \llap{$\mathsurround=0pt#1{#2}$}}

\def\clap#1{\hbox to 0pt{\hss#1\hss}}
\def\mathclap{\mathpalette\mathclapinternal}
\def\mathclapinternal#1#2{%
  \clap{$\mathsurround=0pt#1{#2}$}}

\let\oldstackrel\stackrel
\renewcommand{\stackrel}[2]{\oldstackrel{\mathclap{#1}}{#2}}

\renewcommand{\hbar}{h\mathllap{\overline{\vphantom{h}\hphantom{\rule{4.6pt}{0pt}}}\mspace{0.77mu}}}

\catcode`~=11 % make LaTeX treat tilde (~) like a normal character
\newcommand{\urltilde}{\kern -.06em\lower -.06em\hbox{~}\kern .02em}
\catcode`~=13 % revert back to treating tilde (~) as an active character

%% file: mathlig.tex
\catcode`@=11
\chardef\mathlig@atcode\count255

\def\actively#1#2{\begingroup\uccode`\~=`#2\relax\uppercase{\endgroup#1~}}
\def\mathlig@gobble{\afterassignment\mathlig@next@cmd\let\mathlig@next= }
\def\mathlig@delim{\mathlig@delim}
\def\mathlig@defcs#1{\expandafter\def\csname#1\endcsname}
\def\mathlig@let@cs#1#2{\expandafter\let\expandafter#1\csname#2\endcsname}
\def\mathlig@appendcs#1#2{\expandafter\edef\csname#1\endcsname{\csname#1\endcsname#2}}

\def\mathlig#1#2{\mathlig@checklig#1\mathlig@end\mathlig@defcs{mathlig@back@#1}{#2}\ignorespaces}

\def\mathlig@checklig#1#2\mathlig@end{%
 \expandafter\ifx\csname mathlig@forw@#1\endcsname\relax
 \expandafter\mathchardef\csname mathlig@back@#1\endcsname=\mathcode`#1%
 \mathcode`#1"8000\actively\def#1{\csname mathlig@look@#1\endcsname}%
 \mathlig@dolig#1\mathlig@delim
\fi
\mathlig@checksuffix#1#2\mathlig@end
}

\def\mathlig@checksuffix#1#2\mathlig@end{%
\ifx\mathlig@delim#2\mathlig@delim\relax\else\mathlig@checksuffix@{#1}#2\mathlig@end\fi
}
\def\mathlig@checksuffix@#1#2#3\mathlig@end{%
\expandafter\ifx\csname mathlig@forw@#1#2\endcsname\relax\mathlig@dosuffix{#1}{#2}\fi
\mathlig@checksuffix{#1#2}#3\mathlig@end
}

\def\mathlig@dosuffix#1#2{%
\mathlig@appendcs{mathlig@toks@#1}{#2}%
\mathlig@dolig{#1}{#2}\mathlig@delim
}

\def\mathlig@dolig#1#2\mathlig@delim{%
 \mathlig@defcs{mathlig@look@#1#2}{%
 \mathlig@let@cs\mathlig@next{mathlig@forw@#1#2}\futurelet\mathlig@next@tok\mathlig@next}%
 \mathlig@defcs{mathlig@forw@#1#2}{%
  \mathlig@let@cs\mathlig@next{mathlig@back@#1#2}%
  \mathlig@let@cs\checker{mathlig@chck@#1#2}%
  \mathlig@let@cs\mathligtoks{mathlig@toks@#1#2}%
  \expandafter\ifx\expandafter\mathlig@delim\mathligtoks\mathlig@delim\relax\else
  \expandafter\checker\mathligtoks\mathlig@delim\fi
  \mathlig@next
 }%
 \mathlig@defcs{mathlig@toks@#1#2}{}%
 \mathlig@defcs{mathlig@chck@#1#2}##1##2\mathlig@delim{%
  \ifx\mathlig@next@tok##1%
   \mathlig@let@cs\mathlig@next@cmd{mathlig@look@#1#2##1}\let\mathlig@next\mathlig@gobble
  \fi 
  \ifx\mathlig@delim##2\mathlig@delim\relax\else
   \csname mathlig@chck@#1#2\endcsname##2\mathlig@delim
  \fi
 }%
 \ifx\mathlig@delim#2\mathlig@delim\else
  \mathlig@defcs{mathlig@back@#1#2}{\csname mathlig@back@#1\endcsname #2}%
 \fi
}%

\catcode`@\mathlig@atcode